\documentclass{ws-ijmpcs}

\begin{document}

\markboth{ Agnieszka Janiuk \& Monika Moscibrodzka}
{Accretion and outflow from a magnetized, neutrino cooled torus
around the gamma ray burst central engine}

%
\catchline{}{}{}{}{}
%

\title{Accretion and outflow from a magnetized, neutrino cooled torus
around the gamma ray burst central engine
}

\author{AGNIESZKA JANIUK}

\address{Center for Theoretical Physics, Polish Academy of Sciences\\
Al. Lotnikow 32/46, 02-668 Warsaw, Poland\\
agnes@cft.edu.pl}

\author{MONIKA MOSCIBRODZKA\footnote{
Present address: Department of Physics, University of Nevada Las Vegas, 4505 South Maryland Parkway, Las Vegas, NV 89154, USA  }}

\address{Department of Physics and Department of Astronomy, University of Illinois at Urbana-Champaign\\
1110 West Green Street, Urbana, IL 61801, USA\\
}

\maketitle

\begin{history}
\received{...}
\revised{...}
\end{history}

\begin{abstract}
Gamma Ray Bursts (GRB) are the extremely energetic transient events,
visible from the most distant parts of the Universe. They are most
likely powered by accretion on the hyper-Eddington rates that proceeds
onto a newly born stellar mass black hole. This central engine gives
rise to the most powerful, high Lorentz factor jets that are
responsible for energetic gamma ray emission.  We investigate
the accretion flow evolution in GRB central engine, using the 2D MHD
simulations in General Relativity. We compute the structure and
evolution of the extremely hot and dense torus accreting onto the
fast spinning black hole, which launches the magnetized jets. We
calculate the chemical structure of the disk and account for neutrino
cooling. Our preliminary runs apply to the short GRB case (remnant
torus accreted after NS-NS or NS-BH merger). We estimate the neutrino
luminosity of such an event for chosen disk and central BH mass. 
\end{abstract}

\ccode{PACS numbers: 11.25.Hf, 123.1K}

\section{Model of the hyperaccreting disk}	

The model computations are based on the 2 dimensional, general
 relativistic MHD code HARM2D, described in Ref.~\refcite{gammie}.
The nuclear equation of state is discussed in detail in Ref.~\refcite{janiuk}.
The goal of our preliminary test calculations is to investigate the 
overall structure of an accretion disk in which nuclear reactions take place and 
the gas looses energy via neutrino cooling.

\subsection{Chemical composition}

We assume the gas to be in beta equilibrium, so that the ratio of
proton to neutron satisfies the balance between forward and backward
nuclear reactions.  We assume neutrino cooling via electron, muon and
tau neutrinos in the plasma opaque to their absorption and scattering.
Neutrinos are formed in the following processes: URCA process,
electron positron pair annihilation, nucleon - nucleon bremsstrahlung,
plasmon decay.

\subsection{Dynamical model}

We solve GRMHD equations using numerical code HARM2D. In
our model, the gas is composed of mixture of $p, n, He, e^- $ and $e^+$, so
the total fluid energy is $\rho=\rho_0 + u=\sum n_i m_i + u_i$, where $\rho_0$
is the total rest mass energy density and $u$ is the total internal energy
density and where $i=p, n, He, e^-, e^+$. The total gas pressure is $P=\sum
P_i$. All particles have a common temperature: $T$ ($P_i= n_i kT $) 
and we assume polytropic equation of state $P=(\gamma-1) u$, with adiabatic index
$\gamma=4/3$. We do not include the radiation and neutrino pressure in the
dynamical calculation, but we do account for a neutrino cooling. After each
timestep, we compute the new gas composition and neutrino cooling rates
$q_{\nu}$ as described in Ref.~\refcite{janiuk}. The total internal energy is
reduced using explicit method with $n$-sub-cycles in each time step.
The cooling function $q_{\nu}$ also 
accounts for neutrino absorption and scattering on free neutrons and protons.
The optical thicknesses for absorptions and scatterings are calculated assuming
approximate disk vertical thickness $H \sim 0.5 r$ and local proton and
neutron number densities. Therefore in the dense disk 
regions the cooling is effectively zero.

\subsection{Initial conditions}

We start the numerical calculation from the equilibrium model based on 
Ref.~\refcite{Moncrief} around a spinning black hole. 
The central black hole mass
is $3 M_{\odot}$ and the total mass of the surrounding gas is about 0.1
$M_{\odot}$. HARM2D code is designed to solve the magnetohydrodynamic
equations in the stationary metric. We assume dimensionless spin of a black
hole to be $a_*=0.98$. For higher disk masses, a fully general relativistic
model including the evolution of the space-time and spin of the black hole
should be considered.

The initial magnetic field is assumed with toroidal component of 
vector potential $A_{\varphi} \sim \rho$
(magnetic filed lines follow the constant density surfaces i.e. they form
loops concentrated around the pressure maximum) and normalized with $\beta=
P_{tot,gas}/P_{mag} = 50$.

\section{Results}

\begin{figure}[pb]
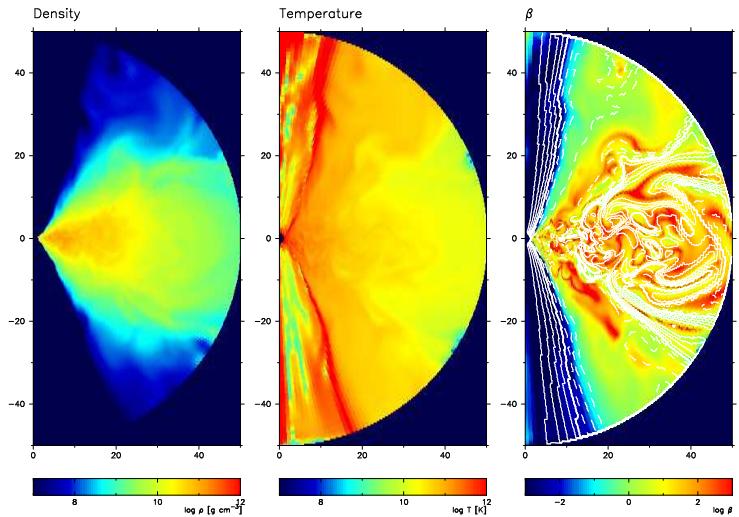

\centerline{
\psfig{file=Janiuk_1_f1.ps,width=6.75cm,angle=270}
\psfig{file=Janiuk_1_f2.ps,width=6.75cm,angle=270}
\psfig{file=Janiuk_1_f3.ps,width=6.75cm,angle=270}
}
\vspace*{8pt}
\caption{Model 1: Structure of an accretion disk in model with no cooling functions taken into
  account in the dynamical evolution. The maps show: (i) density, (ii)
  temperature of the plasma, and (iii) ratio of gas to magnetic pressure, with
  field lines topology (from left to right). }\label{fig1}
\end{figure}

The torus around the spinning black hole at hyper-Eddington  rates is cooled by neutrinos (e.g., as shown in 1-D model in Ref.~\refcite{chen}).
Here we also study such tori, however 
to quantify the effect of neutrino cooling in MHD simulations, 
we run a starting test model with no cooling assumed. 
 In Figure~\ref{fig1}, we show the structure of such a torus (Model 1). 
The snapshots present the baryon 
density ($\rho_0=n_pm_p + 
n_n m_n + n_{He} m_{He}$), gas temperature T and magnetic $\beta$ parameter.
We also show the topology of magnetic field lines. 

A turbulent and dense rotating torus forms in the equatorial plane and extends
towards the marginally stable orbit of the black hole. 
The polar regions are quickly evacuated
and an electromagnetically dominated jet is formed above the poles of the black
hole, as shown before in Refs.~\refcite{McKinneyGammie2004}-\refcite{McKinney2005}. The mass accretion
rate is $\dot{M}=0.4 M_{\odot}/s$.

\begin{figure}[pb]
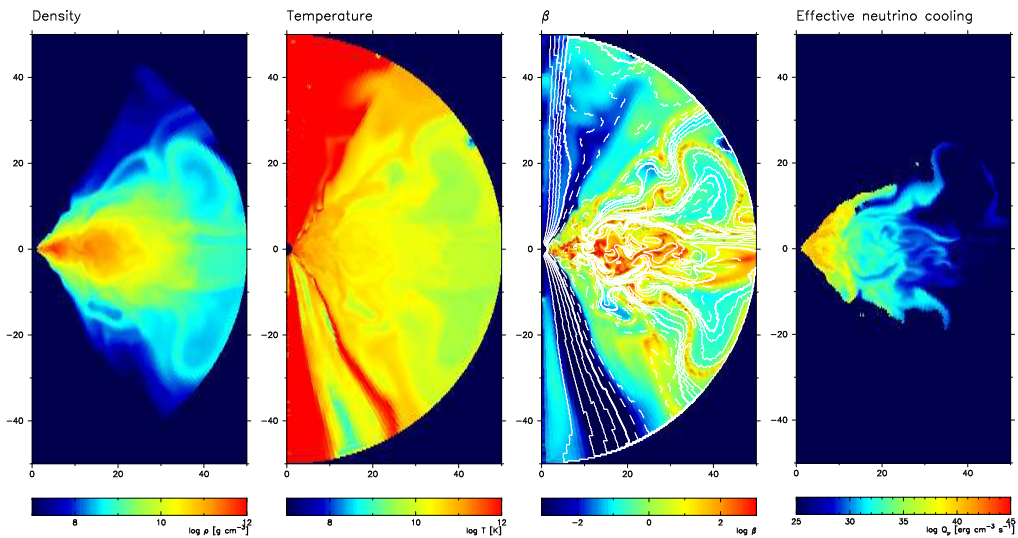

\centerline{
\psfig{file=Janiuk_1_f4.ps,width=7cm,angle=270}
\psfig{file=Janiuk_1_f5.ps,width=7cm,angle=270}
\psfig{file=Janiuk_1_f6.ps,width=7cm,angle=270}
\psfig{file=Janiuk_1_f7.ps,width=7cm,angle=270}
}
\vspace*{8pt}
\caption{Model 2: Structure of accretion disk in model with neutrino cooling taken into
  account in the dynamical evolution.  The maps show: (i) density, (ii)
  temperature of the plasma, (iii) ratio of gas to magnetic pressure, with
  field lines topology, and (iv) the effective neutrino cooling $Q_{\nu}$ (from left to right). Parameters: black hole mass $M =
  3M_{\odot}$, spin $a=0.98$, initial magnetic field normalization $\beta=50$,
  and initial disk mass $M_{\rm disk}= 0.1 M_{\odot}$. The snapshot is at
  t=0.02 s since the formation of the black hole.}\label{fig2}
\end{figure}

In Figure~\ref{fig2} we show Model 2, in which the
neutrino cooling is accounted for in the dynamical evolution. 
The initial conditions in Model 2 are the same as in Model 1. 
The mass accretion rate remains the same as in
Model 1, but the structure of the disk changes. The disk is geometrically
thinner and strongly magnetized in comparison to Model 1. 

We compute the neutrino emissivity at t=0.02 s. The total neutrino luminosity
corrected for absorption and scattering processes, is $L_{\nu}=3\times
10^{54}$ erg s$^{-1}$ (Figure~\ref{fig2}, left panel). This 
number is comparable to
numbers obtained from relativistic hydrodynamical simulations, e.g. 
in Refs.~\refcite{Jaroszynski1993}-\refcite{Birkl2007}. 
Also, the relativistic 
MHD simulations in Ref.~\refcite{Shibata2007}
 reported the neutrino luminosity on the order of $L_{\nu} \lesssim
10^{54}$ erg s$^{-1}$, depending on black hole spin ($a \le 0.9$) 
and torus mass. To compute the
electromagnetic luminosity of the observed GRBs, 
one needs to consider the efficiency of neutrino-antineutrino 
annihilation process, as well as swallowing of some fraction of neutrinos by the black hole due to the curvature effects.

In Figure~\ref{fig2} (most right panel), we show the effective
neutrino cooling. The most neutrinos are formed within $10 R_g$.
We do not see a clear neutrinosphere (a surface at which the optical
thickness for neutrino absorption and scattering is 1.0). Its presence 
as well
as the effect of neutrino trapping will be accounted when
the neutrino pressure is incorporated in the dynamical calculations.

\section*{Acknowledgments}

This research was supported in part by grant NN 203 512638 
from the Polish Ministry of Science and Higher Education.


\end{document}